\documentclass[12pt,preprint]{aastex}
\begin{document}

\title{Discovery of Water Maser Emission in Five AGN and a Possible Correlation
Between Water Maser and Nuclear $2-10$\,keV Luminosities}

\author{Paul T. Kondratko, Lincoln J. Greenhill, James M. Moran}

\affil{Harvard-Smithsonian Center for Astrophysics, 60 Garden St., Cambridge, MA
02138, USA}

\email{pkondrat@cfa.harvard.edu}

\keywords{galaxies: active --- galaxies: individual (2MASXJ08362280+3327383,
NGC6264, UGC09618NED02, IRAS03355+0104, SBS0927+493, MRK0034, NGC3393, NGC5495,
VIIZW073, IC0184, AM2158-380NED02)
--- galaxies: Seyfert --- ISM: molecules --- ISM: jets and outflows --- masers}

\begin{abstract}
We report the discovery of water maser emission in five active galactic nuclei
(AGN) with the 100-m Green Bank Telescope (GBT). The positions of the newly
discovered masers, measured with the VLA, are consistent with the optical
positions of the host nuclei to within $1\sigma$ ($0\rlap{.}''3$ radio and
$1\rlap{.}''3$ optical) and most likely mark the locations of the embedded
central engines. The spectra of three sources, 2MASX\,J08362280+3327383,
NGC\,6264, and UGC\,09618\,NED02, display the characteristic spectral signature
of emission from an edge-on accretion disk with maximum orbital velocity of $\sim
700$\,km\,s$^{-1}$, $\sim 800$\,km\,s$^{-1}$, and $\sim 1300$\,km\,s$^{-1}$,
respectively. We also present a GBT spectrum of a previously known source
MRK\,0034 and interpret the narrow Doppler components reported here as indirect
evidence that the emission originates in an edge-on accretion disk with orbital
velocity of $\sim500$\,km\,s$^{-1}$. We obtained a detection rate of $12\%$\,($5$
out of $41$) among Seyfert\,$2$ and LINER systems with
$10000$\,km\,s$^{-1}<v_{sys}< 15000$\,km\,s$^{-1}$. For the $30$ nuclear water
masers with available hard X-ray data, we report a possible relationship between
unabsorbed X-ray luminosity ($2-10$\,keV) and total isotropic water maser
luminosity, $L_{2-10}\propto L_{\tiny\mbox{H$_2$O}}^{0.5\pm0.1}$, consistent with
the model proposed by Neufeld and Maloney in which X-ray irradiation and heating
of molecular accretion disk gas by the central engine excites the maser emission.
\end{abstract}

\section{Introduction}
\label{introduction} Water maser emission ($\lambda=1.3$ cm) is currently the
only resolvable tracer of warm dense molecular gas in the inner parsec of active
galactic nuclei (AGN) and has been detected to date in approximately $60$ nuclei,
the great majority of which are classified optically as Seyfert\,$2$ or LINER
(e.g., Braatz et al. 1996, 2004; Greenhill et al. 2003; Henkel et al. 2005;
Kondratko et al. 2006; Zhang et al.
2006)\nocite{Braatz1996,Greenhill2003survey,Braatz2004,Henkel2005,Kondratko2006,Zhang2006}.
Because of the association of maser emission with nuclear activity, X-ray
irradiation of molecular gas by the central engine provides the most likely model
for exciting the maser emission \citep*[e.g.,][]{Neufeld1994}. Maser emission
might be associated with Seyfert\,$2$ systems in particular because, over a range
of AGN luminosity, the shielding column density that provides the obscuring
geometry in type\,2 AGN \citep[e.g.,][]{Lawrence1982, Antonucci1993} maintains
not only a reservoir of molecular gas but also physical conditions conducive to
maser action, which are temperatures of $250-1000$\,K and H$_2$ number densities
of $10^{8-10}$\,cm$^{-3}$ \citep*[e.g.,][]{Desch1998}. The importance of the
obscuring geometry in this context is supported by an empirical observation that
water maser sources are found preferentially in nuclei with large hydrogen column
densities \citep[$N_H>10^{24}$\,cm$^{-2}$;][]{Braatz1997,Madejski2006,
Zhang2006}. There is good evidence that LINERs are low-luminosity analogues of
Seyfert\,$2$ systems \citep[e.g.,][]{Ho1997,Ho1999,Ho2003}, which might explain
the association of maser emission with the former. If X-ray irradiation indeed
excites the maser emission \citep[e.g.,][]{Neufeld1994}, then a relationship
between X-ray and water maser luminosities might reflect this dependance.
\cite{Braatz1997} observed no correlation between the two, but that study was
based on only seven water maser systems and relied on the relatively coarse X-ray
data from EXOSAT, GINGA, and ASCA telescopes, with which luminosities can be
difficult to estimate accurately when column densities are large. On the other
hand, \cite{Henkel2005} reported a correlation between infrared and total
isotropic water maser luminosities (i.e., assuming the isotropic emission of
maser radiation), and this might be indirectly indicative of a physical
relationship between X-ray and maser luminosities, mediated by dust reprocessing
(e.g., Franceschini et al. 2005)\nocite{Franceschini2005}. The substantial
increase in the number of known water maser sources over the past few years
(e.g., Greenhill et al. 2003; Braatz et al. 2004; Kondratko et al.
2006)\nocite{Kondratko2006}\nocite{Greenhill2003survey}\nocite{Braatz2004}
combined with the recent growth in the number of AGN for which high-quality X-ray
spectra have been obtained (with angular apertures that isolate the central
engines reasonably well), have enabled a new look at the possibility of a
relationship between X-ray and water maser luminosities.

Very Long Baseline Interferometry (VLBI) maps of seven water maser sources have
been interpreted in a context of a model in which the maser emission traces a
nearly edge-on disk of molecular material $0.1$ to $1$ pc from a supermassive
black hole: NGC\,4258 \citep{Miyoshi1995}, NGC\,1386 \citep{Braatz1997AAS},
NGC\,4945 \citep{Greenhill1997}, NGC\,1068 \citep{Greenhill1997c}, NGC\,3079
\citep{Trotter1998}, IC\,2560 \citep{Ishihara2001}, and Circinus (Greenhill et
al. 2003). As a consequence of these studies, it is believed that maser emission
is detected preferentially in edge-on parsec-scale disks along the diameter
perpendicular to the line of sight (a.k.a. the midline) and close to the
line-of-sight towards the center. These are the loci within disks where the
gradient in line-of-sight velocity is zero and, as a result, the coherent paths
for maser emission are maximized. Thereby, a characteristic spectral signature of
emission from an edge-on disk consists of a spectral complex in the vicinity of
the systemic velocity (low-velocity emission) and two spectral complexes
(high-velocity emission) more or less symmetrically offset from the systemic
velocity by the orbital velocity ($>100$\,km\,s$^{-1}$, based on aforementioned
VLBI studies). Sources that display such spectra constitute approximately $40\%$
of the known nuclear water masers and are referred to here as high-velocity
systems.

Discovery of new high-velocity water maser systems is important because VLBI maps
of these sources can be used to determine pc-scale accretion disk structures, to
estimate accurately black hole masses (e.g., Greenhill \& Gwinn 1997; Greenhill
et al. 2003; Herrnstein et al.
2005)\nocite{Greenhill1997c}\nocite{Greenhill2003}\nocite{Herrnstein2005}, and to
obtain distances to the host galaxies independent of standard candles (Herrnstein
et al. 1999)\nocite{Herrnstein1999}, the latter in cases where a robust disk
model is combined with a measurement of either maser proper motions or drifts in
the line-of-sight velocities of spectral features (i.e., centripetal
acceleration). However, surveys for water maser emission are challenging and
require high sensitivity, wide bandwidth, and high spectral resolution because 1)
the water maser emission is weak ($\ll 1$ Jy), 2) its velocity extent is
determined by the rotational velocity of the accretion disk, which can exceed
$1000$\,km\,s$^{-1}$ and is not known in advance, and 3) the maser lines are
typically narrow ($\le 1$ km s$^{-1}$). The Green Bank Telescope (GBT) of the
NRAO\footnote{The National Radio Astronomy Observatory is operated by Associated
Universities, Inc., under cooperative agreement with the National Science
Foundation} and its wide-bandwidth spectrometer together constitute the most
sensitive observing system currently in operation at $\lambda=1.3$\,cm; as a
result, detection rates for samples with comparable distributions of distance are
highest for surveys conducted with the GBT. For instance, the seminal survey of
AGN with $1\sigma\sim60$\,mJy sensitivity (in $\sim1$\,km\,s$^{-1}$ spectral
channels) by \cite{Braatz1996} and a search with $1\sigma\sim14$\,mJy sensitivity
(in $1.3$\,km\,s$^{-1}$ spectral channels) with the $70$-m NASA Deep Space
Network (DSN) antennas (Greenhill et al. 2003; Kondratko et al.
2006)\nocite{Greenhill2003survey}\nocite{Kondratko2006} yielded incidence rates
of water maser emission among nearby ($v_{sys}<7000$\,km\,s$^{-1}$) Seyfert\,2
and LINER systems of $\sim7\%$ and $\sim10\%$, respectively. However, a GBT
survey conducted with $1\sim3$\,mJy sensitivity (converted to $1.3$\,km\,s$^{-1}$
spectral channels) resulted in an incidence rate of $\sim20\%$ among Seyfert\,$2$
and LINER systems with $v_{sys}<7500$\,km\,s$^{-1}$ \citep{Braatz2004}.

We used the GBT to survey $56$\,AGN with $10000 < v_{sys} < 30000$\,km\,s$^{-1}$
selected from the NASA Extragalactic Database (NED). Previous surveys have
concentrated on detection of water maser emission mostly among nearby sources
\citep[cf.][]{Henkel1998,Barvainis2005}. In particular, galaxies with
$v_{sys}>10000$\,km\,s$^{-1}$ constitute only $\sim7\%$ and $14\%$ of AGN in the
two largest surveys to date (Braatz et al. 1996 and Kondratko et al. 2006,
respectively)\nocite{Braatz1996} and, as a result, only $4$ out of $\sim60$ known
maser sources lie beyond $10000$\,km\,s$^{-1}$ (Tarchi et al. 2003; Henkel et al.
2005; Kondratko et al. 2006, Zhang et al. 2006)\nocite{Kondratko2006, Henkel2005,
Tarchi2003, Zhang2006}. We have discovered water maser emission in five AGN
between $10000$ and $15000$\,km\,s$^{-1}$. In this paper, Section \ref{results},
we present spectra and sub-arcsecond positions for the detected emission and
address survey statistics. In Section \ref{discussion}, we discuss a possible
correlation between X-ray and water maser luminosities.

\section{Observations}

The survey was conducted during the 2003-2004 northern winter with the GBT using
the observing system and correlator setup identical to that described in
\cite{Braatz2004}. The channel spacing was $24.4\,\mbox{kHz}$ and the
instantaneous bandwidth was $380\,\mbox{MHz}$ ($\approx 5600$\,km\,s$^{-1}$ for a
representative recessional velocity of $12500$\,km\,s$^{-1}$, assuming the
optical definition of Doppler shift). To obtain total-power spectra of each
source, we nodded the telescope by $3'$ every $2$\,min between two positions on
the sky so that each target was always present in one of the two GBT beams for
each polarization. System temperatures were measured against a calibrated noise
source injected at the receiver and ranged from $28$ to $59$\,K depending on
elevation and weather conditions. By comparing maser line amplitudes among beams
and polarizations, we estimate that the calibration of the system temperature is
accurate to within $30\%$. This uncertainty dominates the error budget for the
flux density scale in the survey. Antenna pointing corrections
--- obtained every $\sim30$ minutes using sources from the VLA Calibrator
Manual --- were typically $<6''$, which corresponds to a $<8\%$ loss in source
flux density for a $36''$ beamwidth (FWHM) at $1.3$\,cm. For the observations
reported here, the wind speed at the GBT site was generally $<13$\,mph, which
corresponds to a one-dimensional rms tracking error of $<5''$ and a signal loss
of $<6\%$ (with the exception of CGCG\,482-008 and MRK\,0948, which were observed
with wind gusts up to $20$\,mph, corresponding to a one-dimensional rms tracking
error of $<13''$ and a signal loss of $<29\%$; Condon 2003)\nocite{Condon2003}.
Data were reduced using scripts written in the Interactive Data Language. We
subtracted a running boxcar average of width $6.25$\,MHz to remove systematic
baseline structure from the total-power spectra. To convert the spectra to flux
density units, we used the gain curve obtained by the GBT staff based on
measurements of opacity corrected antenna temperature for NGC\,7027 at
$\sim1.4$\,cm (R. Maddalena 2003, private communication). The resulting
$1\,\sigma$ noise levels attained in an integration of $\sim30$\,minutes total
and corrected for atmospheric opacity estimated from tipping scans (typically
$\sim0.06$) were $3-6$\,mJy in a $24.4\,\mbox{kHz}$ channel. The spectra
presented here have been iteratively Hanning smoothed to an effective resolution
of $108\,\mbox{kHz}$.

\section{Results}
\label{results} 
\begin{figure*}[!h]
\epsscale{0.85} \plotone{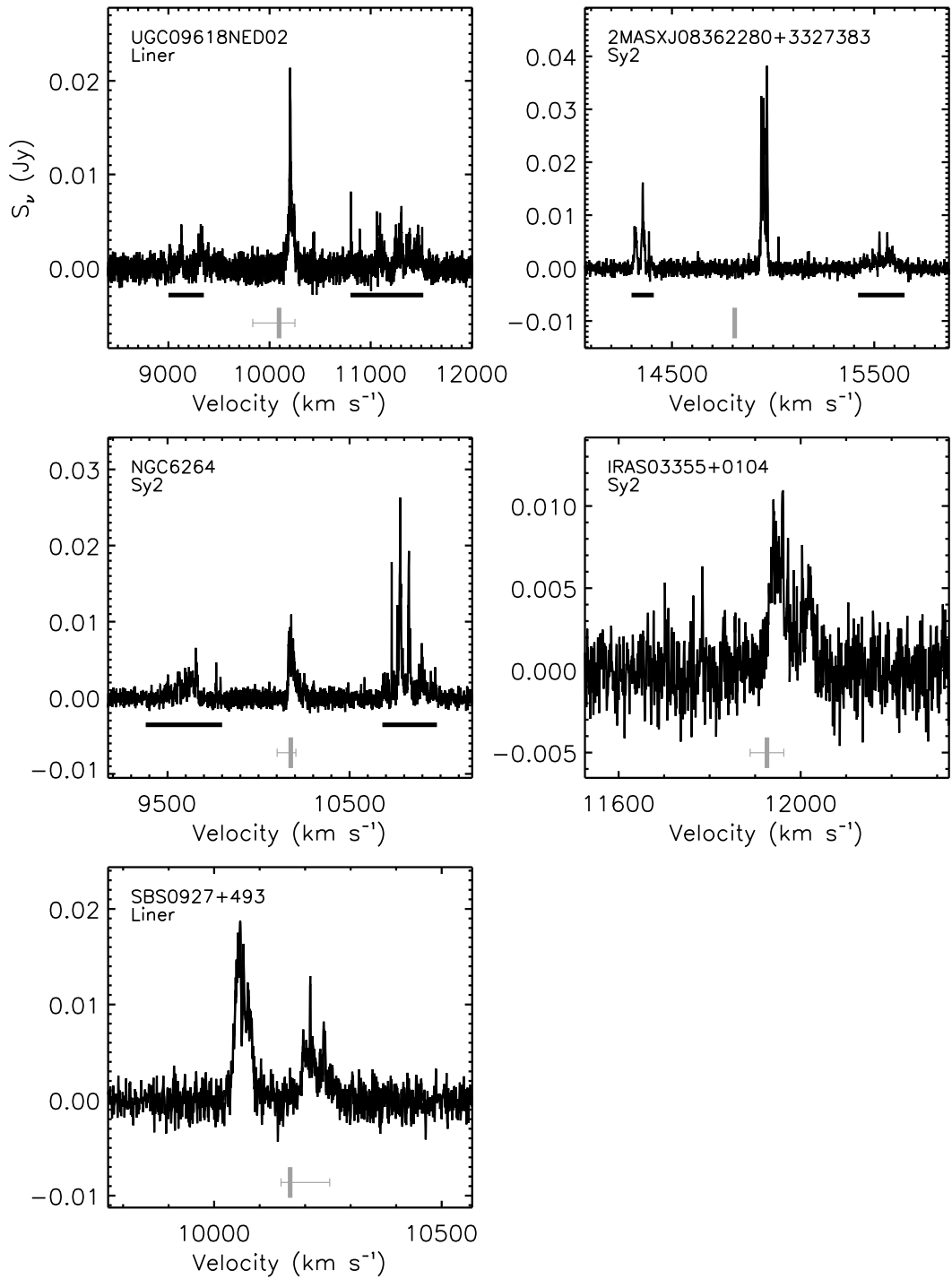} \caption{Spectra of UGC\,09618\,NED02,
2MASX\,J08362280+3327383, NGC\,6264, IRAS\,03355+0104, SBS\,0927+493 obtained
with the Green Bank Telescope. Vertical bars mark the optical heliocentric
systemic velocities of the host galaxies listed in the NED, while the associated
error bars show the range of listed systemic velocity estimates. Horizontal bars
indicate approximate velocity ranges of high-velocity emission. To illustrate the
difference in the velocity scale among spectra, the tick marks are placed every
$100$\,km\,s$^{-1}$ for each x-axis. \label{figure}}
        \hrulefill\
\end{figure*}

In a survey with the GBT of $56$\,AGN with
$10000$\,km\,s$^{-1}<v_{sys}< 30000$\,km\,s$^{-1}$ selected from the NED (Table
\ref{Survey}), we have detected five new water maser sources: UGC\,09618\,NED02,
2MASX\,J08362280+3327383, NGC\,6264, IRAS\,03355+0104, SBS\,0927+493 (Fig.
\ref{figure} and Table \ref{Table}). The five discoveries were subsequently
confirmed with the Very Large Array (VLA) using narrow observing bandwidths
($6.25-12.5$\,MHz). In addition, the UGC\,09618\,NED02, 2MASX\,J08362280+3327383,
NGC\,6264, and SBS\,0927+493 broadband (i.e., $350$\,MHz) maser spectra were
confirmed by at least one observation on a different day with the GBT. The nuclei
that are host to the detected maser emission are spectroscopically classified as
Seyfert\,$2$ or LINER (Table \ref{Table}). Positions of the maser emission
measured with the VLA are consistent with optical positions of the AGN to within
$1\sigma$ (typically $0\rlap{.}''3$ radio and $1\rlap{.}''3$ optical, or of order
$1$\,kpc at a recessional velocity of $15000$\,km\,s$^{-1}$), which is suggestive
of an association of the detected emission with nuclear activity.

The maser spectra of UGC\,09618\,NED02, 2MASX\,J08362280+3327383, and NGC\,6264
show a characteristic spectral signature of emission from an edge-on disk: two
high-velocity complexes approximately symmetrically offset from the systemic
velocity and a third spectral complex in the vicinity of the systemic velocity.
If the highly red- and blue-shifted emission in these systems indeed represents
high-velocity emission, then the maximum orbital velocities of the disks as
traced by the maser emission are $\sim 1300$\,km\,s$^{-1}$,
$\sim700$\,km\,s$^{-1}$, and $\sim800$\,km\,s$^{-1}$, respectively, making
UGC\,09618\,NED02 the fastest known rotator. (The previously fastest rotator was
NGC\,4258 at $\sim 1200$\,km\,s$^{-1}$; Modjaz et al. 2005.)\nocite{Modjaz2005}
Because the high-velocity emission extends over $\Delta v\sim 750$\,km\,s$^{-1}$
for UGC\,09618\,NED02, it must occupy a fractional range of radii $\Delta r/R
\approx 2\Delta v/v \sim 1.6$ assuming Keplerian rotation (where we have
differentiated the Keplerian rotation law, $v\propto r^{-0.5}$, with respect to
$r$). For a range in disk sizes from that of NGC\,4258 ($0.16-0.28$\,pc) to
NGC\,1068 ($0.6-1.1$\,pc), the maser emission in this source traces a disk of
radial extent $\Delta r\sim 0.36-1.4$\,pc. The corresponding central mass would
be $(6-24)\times 10^7\,M_{\odot}$, which is larger than black hole masses
measured with VLBI to date \citep[the highest is NGC\,4258 at
$3.9\times10^7\,M_{\odot}$][]{Herrnstein2005}. The anticipated centripetal
acceleration
--- that is the secular velocity drift of the systemic feature --- would be
$0.3-11$\,km\,s$^{-1}$\,yr$^{-1}$, which should be readily detectable within one
year using single dish monitoring. The remaining two detections ---
IRAS\,03355+0104 and SBS\,0927+493
--- display spectral feature(s) only in the vicinity of the systemic velocity of the
host galaxy, which makes physical interpretation difficult.

\begin{figure*}[!h]
\epsscale{0.85} \plotone{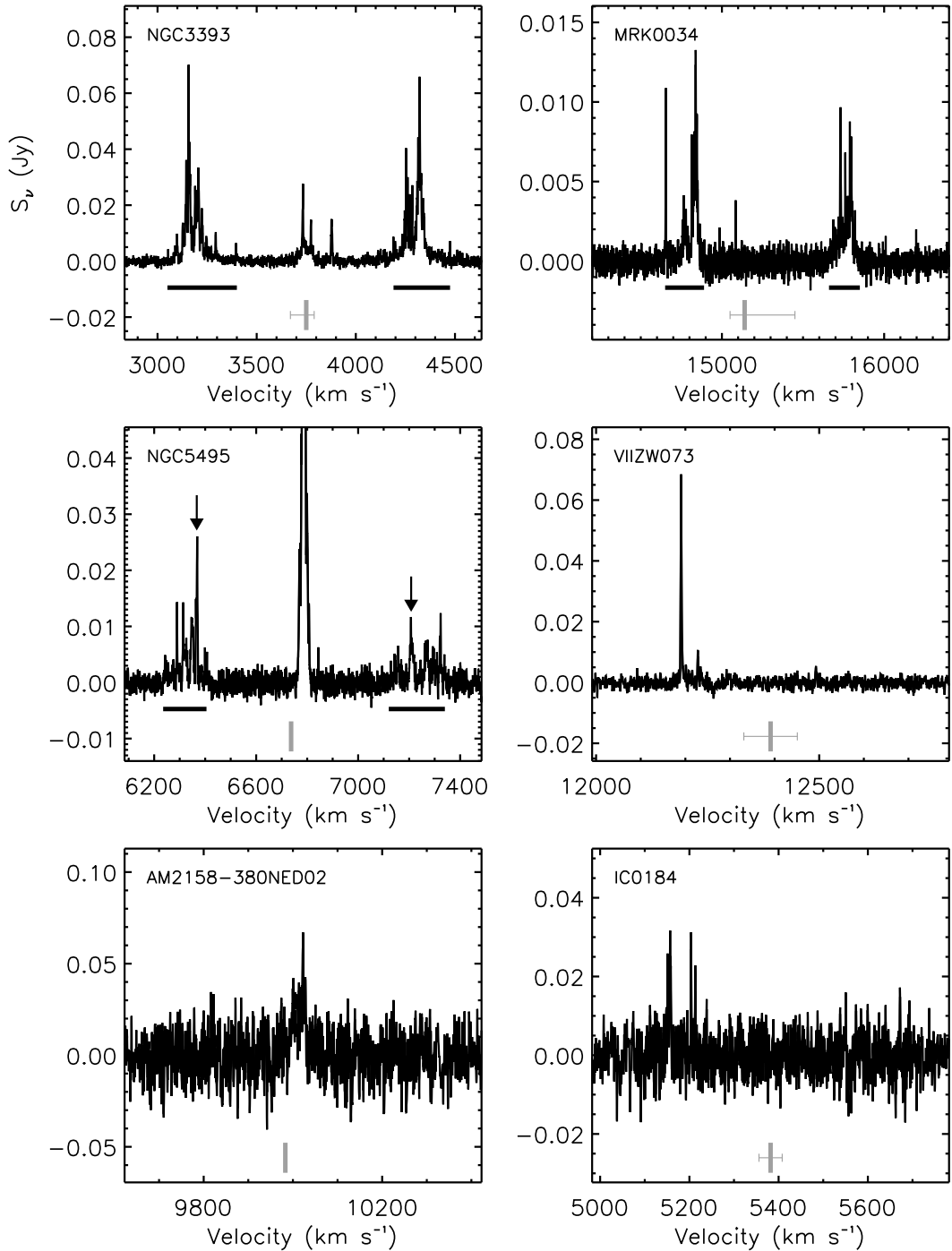} \caption{Spectra of previously known masers in
NGC\,3393, MRK\,0034, NGC\,5495, VII\,ZW\,073, AM\,2158-380\,NED02, and IC\,0184
obtained with the Green Bank Telescope. The arrows indicate NGC\,5495
high-velocity features marginally detected with the DSN \citep{Kondratko2006}.
Vertical bars mark the optical heliocentric systemic velocities of the host
galaxies listed in the NED, while the associated error bars show the range of
listed systemic velocity estimates. Horizontal bars indicate approximate velocity
ranges of high-velocity emission. To illustrate the difference in the velocity
scale among spectra, the tick marks are placed every $100$\,km\,s$^{-1}$ for each
x-axis. \label{other}}
        \hrulefill\
\end{figure*}

We also used the GBT to obtain high signal-to-noise-ratio (SNR) spectra of known
sources, including those recently discovered with the DSN (cf. Henkel et al.
2005; Kondratko et al. 2006; Fig. \ref{other} and Table \ref{TableOther}).
Spectra of MRK\,0034, NGC\,3393, NGC\,5495, VII\,ZW\,073, and IC\,0184 presented
here are the most sensitive obtained to date (cf. Henkel et al. 2005; Kondratko
et al. 2006)\nocite{Henkel2005}\nocite{Kondratko2006}. The narrow Doppler
components in the spectrum of MRK\,0034 reported here were not evident in the
previous spectra of the source obtained with $4.65$\,km\,s$^{-1}$ channels.
Absent narrow lines, \cite{Henkel2005} suggested the possibility that the maser
emission in MRK\,0034 might be excited by jet activity, which has been associated
with broad spectral features in other AGN \citep[e.g.,][]{Claussen1998,
Gallimore2001, Peck2003}. However, the new spectrum adds support for the
hypothesis that MRK\,0034 is a high-velocity system. In this context and
considering the large uncertainty on the recessional velocity of the galaxy (see
Fig. \ref{other}), the two spectral complexes are consistent with blue- and
red-shifted high-velocity emission that is symmetrically displaced from the
systemic velocity by $\sim500$\,km\,s$^{-1}$. The GBT spectrum of NGC\,3393 shows
a new systemic complex at $\sim 3750$\,km\,s$^{-1}$ as well as weak
($\sim5$\,mJy) high-velocity lines at $\sim3098$, $\sim3293$, $\sim3396$,
$\sim4190$, and $\sim4475$\,km\,s$^{-1}$ (Fig. \ref{other}). We also confirmed
the low-velocity line at $\sim3878$\,km\,s$^{-1}$ in the NGC\,3393 spectrum and
the high-velocity lines in the NGC\,5495 spectrum, which were all marginally
detected with the DSN (Kondratko et al. 2006).

Our detection rate is consistent with the survey being sensitivity limited. Among
Seyfert\,$2$ and LINER systems with $10000$\,km\,s$^{-1}<v_{sys}<
15000$\,km\,s$^{-1}$, we obtain a detection rate of $5$ out of $41$ or $12\%\pm
5\%$, which should be compared to the incidence rates of $20\%\pm5\%$ ($15$ out
of $75$) and $14\%\pm5\%$ ($7$ out of $49$) in velocity bins $0-5000$ and
$5000-12000$\,km\,s$^{-1}$, respectively, from a GBT survey of the same
sensitivity (Braatz et al. 2004; to estimate the errors, we used the unbiased
maximum likelihood estimator for the variance of the Binomial parameter
$p$)\nocite{Braatz2004}. The detection rate decreases with distance because the
spectral rms noise level was approximately the same for each source in both
surveys. The detection rate of high-velocity emission is $60\%\pm 24\%$ ($3$ out
of $5$) in this survey or $44\%\pm 13\%$ ($7$ out of $16$) if we include the GBT
detections reported in Braatz et al. (2004; where we excluded the starburst
galaxy NGC\,2782)\nocite{Braatz2004}. These rates should be compared to the
analogous rate of $27\%\pm 12\%$ ($4$ out of $15$) in the DSN survey (Greenhill
et al. 2003; Kondratko et al.
2006)\nocite{Greenhill2003survey}\nocite{Kondratko2006}. Although the
uncertainties due to counting statistics are large, the higher detection rates
for high-velocity emission in the GBT surveys may have been a consequence of
higher sensitivity, and perhaps finer spectral resolution as well.

\section{Discussion}
\label{discussion} 
\begin{figure*}[!h]
\epsscale{0.85} \plotone{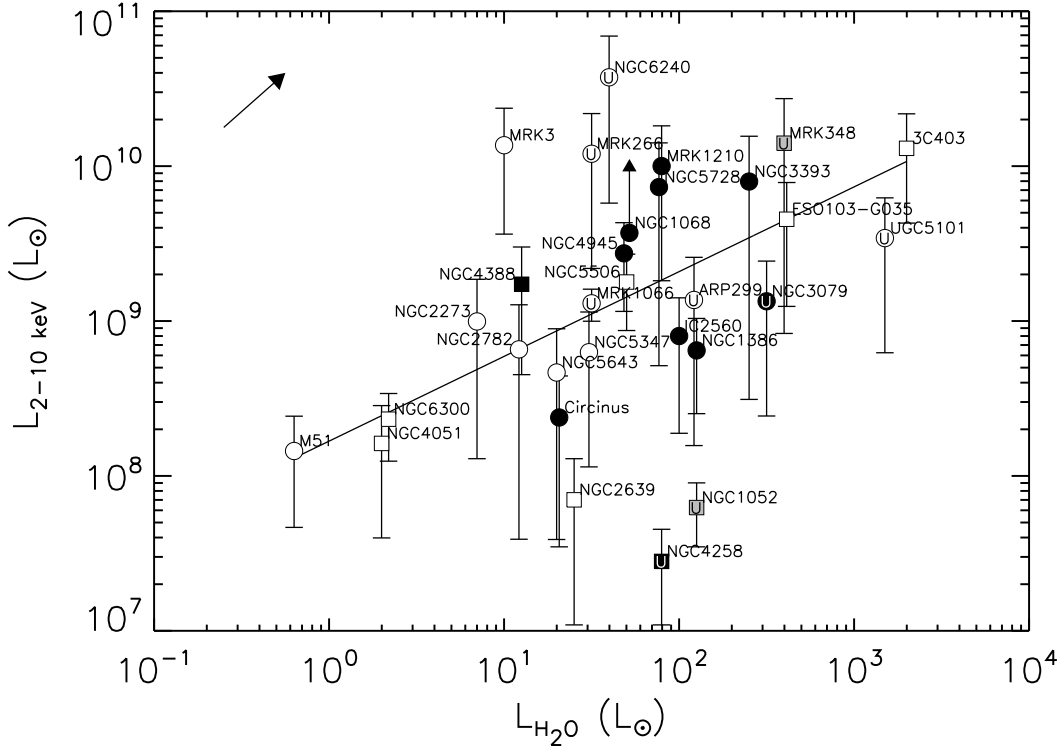} \caption{Unabsorbed X-ray luminosity
($2-10$\,keV) versus total isotropic water maser luminosity where the data points
are from Table \ref{XrayTable}. Symbol shading discriminates among maser types:
{\it(black)} high-velocity systems (see definition in Section
\ref{introduction}), {\it(open)} sources that are not known to be high-velocity
systems, and {\it(grey)} maser sources known to be associated with radio jets
rather than disks. Symbol shape differentiates between Compton-thick {\it(round)}
and Compton-thin {\it(square)} objects. Galaxies with multiple nuclei, AGN with
maser emission tied to jet activity, and LINER systems (see the text) are
labelled by a letter U. Vertical error bars indicate the range of luminosity
estimates given in the literature for independent data sets (Table
\ref{XrayTable}), and the symbols are placed at the midpoints. The arrow in the
top left corner illustrates the effect of a hypothetical factor of $1.5$ increase
in distance (i.e., points move up and to the right by this amount). For the full
sample with no LINERs, systems with multiple nuclei, and ``jet masers," we obtain
a correlation coefficient of $0.5\pm 0.1$ and a slope of $0.5\pm0.1$ (black line
shows the results of a linear fit), where the asymmetric uncertainties in
ordinate values have been taken into account (see Table \ref{fits}). The lower
limit on X-ray luminosity of NGC\,1068 was not used in the fit. \label{Xray}}
        \hrulefill\
\end{figure*}

Using the sample of known nuclear masers with available
$>2$\,keV X-ray data, we have identified a possible relationship between
unabsorbed X-ray luminosity ($2-10$\,keV) and integrated water maser luminosity,
assuming isotropic emission of radiation. In log-log space (Fig. \ref{Xray} and
Table \ref{XrayTable}), we obtain a slope of $m=0.4\pm 0.1$ and a Spearman
correlation coefficient of $\rho=0.4\pm 0.1$ (Table \ref{fits}). The probability
that a random data set of the same size would yield a larger magnitude of the
correlation coefficient (hereafter referred to as the significance level) is
$0.07\pm0.07$.

Under the assumption of a thin viscous accretion disk obliquely illuminated at an
angle $\cos^{-1}\mu$ by a central X-ray source, the outer radius at which the
disk becomes atomic and the maser emission ceases is given by $R_{cr}\propto
L_{2-10}^{-0.43}(\dot{m}/\alpha)^{0.81}\mu^{-0.38}M_{BH}^{0.62}$
\citep{Neufeld1995}, where $L_{2-10}$ is the $2-10$\,keV luminosity, $M_{BH}$ is
the mass of the central black hole, $\dot{m}$ is the mass accretion rate, and
$\alpha$ is the standard dimensionless viscosity parameter \citep{Shakura1973}.
If we assume that X-ray heated accretion disks emit water maser radiation with a
surface luminosity of $\sim10^{2\pm 0.5}\,L_{\odot}\,\mbox{pc}^{-2}$ \citep[and
hence $L_{\mbox{\tiny H$_2$O}} \propto R^2_{cr}$;][]{Neufeld1994}, that
$2-10$\,keV luminosity is a fraction $\gamma$ of AGN bolometric luminosity
($L_{2-10}=\gamma\,L_{Bol}$), that central engines radiate with an Eddington
ratio $\eta$ ($L_{Bol}= \eta L_{Edd} \propto \eta M_{BH}$), and that they convert
rest mass to energy with an accretion efficiency $\epsilon=L_{Bol}/\dot{m}c^2$
\citep*[$\epsilon\sim 0.1$;][]{Frank2002}, then we obtain $L_{2-10}\propto
L_{\mbox{\tiny
H$_2$O}}^{0.5}\,[(\epsilon\alpha)^{0.81}\gamma^{1.4}\mu^{0.38}\eta^{0.62}]$, in
agreement with the observed trend.

The scatter about the linear relationship in Fig. \ref{Xray} in excess of the
reported uncertainties could be attributed to differences in the parameters
$\alpha$, $\epsilon$, $\gamma$, $\mu$, and $\eta$ from source to source. For
instance, although $0.01\lesssim \gamma \lesssim 0.03$ based on an average quasar
spectral energy distribution \citep[SED;][]{Fabian1999, Elvis2002}, the actual
variation in $\gamma$ among individual Seyfert systems is more than an order of
magnitude (e.g., $0.01<\gamma<0.6$; Kuraszkiewicz et al.
2003)\nocite{Kuraszkiewicz2003}. Furthermore, although $\eta$ is relatively
constrained for Seyfert nuclei \citep[e.g., $0.01\lesssim\eta \lesssim
1$;][]{Padovani1989, Wandel1999, Satyapal2005}, it might be as low as $10^{-6.7}$
for LINER systems \citep[e.g.,][]{Satyapal2005}. For the LINER in NGC\,4258 where
black hole mass is well constrained with VLBI, we obtain
$10^{-3.6}<\eta<10^{-2.9}$, where we adopted $M_{BH}=3.9\times 10^7\,M_{\odot}$
(Herrnstein et al. 1999)\nocite{Herrnstein1999},
$L_{2-10}=10^{40.6-41.2}$\,erg\,s$^{-1}$ (Table \ref{XrayTable}), and
$\gamma=0.03$ as determined from SEDs of $7$ low-luminosity AGNs \citep{Ho1999b}.
Because of the broader range of Eddington ratio specific to LINERs, we have
considered a Seyfert-only sub-sample. Furthermore, we have also removed galaxies
in which the maser sources are associated with jets (for which the Neufeld \&
Maloney model does not apply), and galaxies with multiple nuclei (wherein it may
be uncertain which nucleus is responsible for the maser emission or the
production of maser luminosity may in some way be affected by processes specific
to mergers). The active galaxies with multiple nuclei are MRK\,266
\citep[e.g.,][]{Wang1997}, ARP\,299 \citep[e.g.,][]{Ballo2004}, NGC\,6240
\citep[e.g.,][]{Lira2002}, and MRK\,1066 \citep[e.g.,][]{Gimeno2004}. If these
mergers, known ``jet masers" \citep[NGC\,1052 and MRK\,348;][]{Claussen1998,
Peck2003}, and LINERs (NGC\,3079, NGC\,4258, NGC\,6240, NGC\,1052, UGC\,5101) are
omitted from the sample, then we obtain a slope of $m=0.5\pm 0.1$ and a Spearman
correlation coefficient of $\rho=0.5\pm 0.1$, the latter with a significance
level of $0.03\pm 0.04$ (Table \ref{fits}). To test for the robustness of this
strongest correlation, we removed the data points with lowest {\it and} highest
maser luminosity and obtained $m=0.5\pm 0.1$ and $\rho=0.4\pm 0.1$, the latter
with a significance level of $0.1\pm 0.1$ (Table \ref{fits}).

The appearance of a correlation may be surprising given large uncertainties in
X-ray and maser luminosities. Both are time variable, and it is easy to imagine
some independence in that variability. As well, measured X-ray luminosities for
heavily absorbed sources depend sensitively on the details of instrumentation and
subsequent data modelling, although we have accounted for this to some extent
with the adopted uncertainties. Moreover, isotropic maser luminosities should be
regarded with some caution because maser beam angles are often poorly constrained
and may be quite small due to source geometry and propagation effects
\citep*[e.g.,][]{Deguchi1989, Kartje1999}. For instance, the beam angle for
NGC\,4258 is on the order of $7^\circ$ \citep{Miyoshi1995} and, as a result, the
actual luminosity is a factor of $4\pi/\Omega=4\pi/(2\pi\times 7^\circ)\sim 16$
smaller than isotropic. Accretion disk warping may also bias maser brightness, as
beamed radiation may be directed away from the observer. However, one might argue
conversely based on the putative correlation that beam angles among water maser
systems are relatively similar and variability is a second order effect on
average. Furthermore, since the estimated power law index of $0.5\pm 0.1$ is
consistent with the value ($0.5$) predicted by the \cite{Neufeld1995} model ---
notwithstanding the large uncertainties that enter through the Eddington ratio,
bolometric correction, viscosity, and accretion efficiency --- one might venture
to propose that these parameters are similar among sources, at least to within
factors of a few.

The full sample presented here is inhomogeneous in that it contains AGN that are
Compton-thin and Compton-thick, as well as water maser sources for which there is
evidence of an origin in accretion disk material and those for which available
data do not enable any particular physical classification. We have computed $m$
and $\rho$ for various sub-samples and find ranges of $0.3-0.6$ and $0.1-0.5$,
respectively (Table \ref{fits}). The correlation coefficient is largest
($0.5\pm0.1$) and most statistically significant ($0.03\pm 0.04$) for the sample
absent LINERs, multiple-nuclei systems, and ``jet masers." No correlation is
apparent for high-velocity systems due to the small sub-sample size. We have also
considered Compton-thick and Compton-thin sub-samples. Since a minimum H$_2$
density of $\sim 10^8$\,cm$^{-3}$ is required by the presence of water maser
emission \citep[e.g.,][]{Desch1998}, the column within a disk becomes
Compton-thick for lengths greater than $\sim0.002$\,pc and, consequently, it has
been suggested that the pc-scale accretion disks that host the maser emission
also provide the Compton obscuration (Greenhill et al. 2003; Herrnstein et al.
2005; Fruscione et al.
2005)\nocite{Greenhill2003}\nocite{Herrnstein2005}\nocite{Fruscione2005}. We note
however that there is no theoretical reason why maser emission should not
originate in a Compton-thin environment. Investigation into properties of
astrophysical gas by \cite*{Maloney1996} suggests that fractional water
abundances of $10^{-6}-10^{-4}$ and temperatures of $250-1000$\,K --- both
conditions necessary for maser action
--- are possible for column densities of $\sim10^{23}$\,cm$^{-2}$
(their Fig. 10). Assuming that the edge-on pc-scale accretion disks also provide
the line-of-sight X-ray obscuration, the fact that Compton-thin environments are
conducive to maser action is also supported by an empirical observation that
approximately $33\%$ of the known nuclear masers lie in Compton-thin systems
(Table \ref{XrayTable}) and that several of these are also high-velocity systems
(such as NGC\,4388, NGC\,4258, and NGC\,5728). If pc-scale accretion disks indeed
provide the line-of-sight obscuring column density, factors such as the disk
inclination, warping, and filling factor might affect the X-ray absorption column
and it is thus possible that the intrinsic relationship between water maser and
X-ray luminosities is different for Compton-thick and -thin sources. However, the
two sub-samples are too small to assess any difference (Table \ref{fits}).

\section{Conclusion}

In a survey with the GBT of $56$\,AGN with $10000$\,km\,s$^{-1}<v_{sys}<
30000$\,km\,s$^{-1}$, we have detected five new water maser sources. The spectra
of three sources display the characteristic spectral signature of emission from
an edge-on accretion disk. For the $30$ nuclear water masers with available hard
X-ray data, we have found evidence of a possible relationship between unabsorbed
X-ray luminosity ($2-10$\,keV) and total isotropic water maser luminosity. The
power law index of $0.5\pm0.1$ is consistent with the \cite{Neufeld1995} model in
which X-ray irradiation of molecular accretion disk gas by a central engine
excites maser emission that is most intense in the disk plane. The appearance of
a correlation may be surprising considering the large uncertainties in X-ray and
maser luminosities, at least in part due to source variability and assumption of
isotropic emission of maser radiation. However, one might interpret the
correlation as indirectly supporting the proposition that Eddington ratios,
bolometric corrections, viscosities, accretion efficiencies, and maser beaming
angles are similar among water maser systems, and variability is a second order
effect on average. Evaluation of the putative correlation would be greatly helped
by an increase in the number of known maser systems and reduction in measurement
and systematic uncertainties in the luminosity data. This will require
measurement of X-ray spectra above $10$\,keV, new radio surveys of AGN (in search
of maser emission), and VLBI mapping to quantify accretion disk structure. If
verified and strengthened, the proposed relation between luminosities would be
valuable in the modelling of maser excitation, identification of maser-rich
samples of AGN using hard X-ray sky surveys, and perhaps modelling of the hard
X-ray background in the era of the Square Kilometer Array, which will enable
detection of many thousands of maser galaxies \citep{Morganti2004}.

\section{Acknowledgements}

We thank J.\,Braatz, M.\,Elvis, R.\,Narayan, M. Reid, and B.\,Wilkes for helpful
discussions, C.\,Bignell for flexibility in GBT scheduling, and R.\,Maddalena for
providing us with the $1.3$\,cm gain curve. We thank J.\,Braatz as well for help
in set-up and observing. This research has made extensive use of the NASA/IPAC
Extragalactic Database (NED) which is operated by the Jet Propulsion Laboratory
(JPL), California Institute of Technology, under contract with NASA. This work
was supported by GBT student support program, grants GSSP004-0005 and
GSSP004-0011.

\bibliography{ms}
\clearpage
\clearpage
\begin{deluxetable}{llllccccc}
\tablecaption{Galaxies Surveyed for Water Maser Emission with the Green Bank
Telescope. [Refer to the source file for the complete version of this
table.]\label{Survey}} \tablehead{
     \colhead{Galaxy}              &
     \colhead{Type\tablenotemark{(a)}}              &
     \colhead{$\alpha_{2000}$}      &
     \colhead{$\delta_{2000}$\tablenotemark{(a)}}   &
     \colhead{v$_{sys}$\tablenotemark{(a)}}    &
     \colhead{Date}           &
     \colhead{$T_{sys}$\tablenotemark{(b)}}      &
     \colhead{$1\sigma$\tablenotemark{(c)}} \\
     \colhead{}             &
     \colhead{}             &
     \colhead{(hh~mm~ss)}     &
     \colhead{(dd~mm~ss)}     &
     \colhead{(km s$^{-1}$)}       &
     \colhead{}        &
     \colhead{(K)}          &
     \colhead{(mJy)}
}

\startdata

UM213 & LINER & 00~12~17.83 & $-$00~06~10.6 &         12149 & 11-10-2003 &           30 & 4.6\\
MRK0948 & Sy2 & 00~28~14.34 & $+$07~07~45.4 &         12022 & 11-08-2003 &           42 & 4.8\\
UM254 & Sy2 & 00~31~34.27 & $-$02~09~16.8 &         13311 & 11-10-2003 &           32 & 5.9\\
FGC0061 & Sy2 & 00~34~43.51 & $-$00~02~26.7 &         12610 & 11-10-2003 &           28 & 3.7\\
LEDA093200 & Sy2 & 00~41~35.07 & $-$09~21~52.2 &         14234 & 11-10-2003 &           29 & 4.0\\

\enddata

\tablenotetext{(a)}{Type, position, and heliocentric optical systemic velocity
obtained from the NED at the outset of the survey in 2002.}

\tablenotetext{(b)}{Average system temperature.}

\tablenotetext{(c)}{Rms noise in a $24.4$\,kHz spectral channel, corrected for
atmospheric opacity (typically from $0.02$ to $0.07$) and for the dependence of
antenna gain on elevation.}

\end{deluxetable}

\clearpage
\thispagestyle{empty}

\begin{deluxetable}{llrrrllrrrrr}
\tabletypesize{\footnotesize} \rotate \tablewidth{8.2in} \tablecaption{Newly
Discovered Nuclear Water Maser Sources.\label{Table}} \tablehead{
     \colhead{Galaxy}              &
     \colhead{Type\tablenotemark{(a)}}              &
     \colhead{$\alpha_{2000}$\tablenotemark{(b)}}      &
     \colhead{$\delta_{2000}$\tablenotemark{(b)}}   &
     \colhead{v$_{sys}$\tablenotemark{(c)}}    &
     \colhead{Date}           &
     \colhead{$T_{sys}$\tablenotemark{(d)}}      &
     \colhead{$1\sigma$\tablenotemark{(e)}} &
     \colhead{$\Delta\nu$\tablenotemark{(f)}} &
     \colhead{BW\tablenotemark{(g)}} \\
     \colhead{Name}             &
     \colhead{}             &
     \colhead{(hh~mm~ss)}     &
     \colhead{(dd~mm~ss)}     &
     \colhead{(km s$^{-1}$)} &
     \colhead{}             &
     \colhead{(K)} &
     \colhead{(mJy)}             &
     \colhead{(kHz)} &
     \colhead{(MHz)}

}

\startdata

IRAS\,03355+0104     & Sy2 & 03~38~10.38 & $+$01~14~18.2 & 11926  & 11-08-2003 & 38 & 3.2  & $195$ & $12.5$ \\
                  &       & 03~38~10.38 & $+$01~14~18.3            \\

2MASX\,J08362280+3327383     & Sy2 & 08~36~22.80 & $+$33~27~38.6 & 14810  & 01-15-2005 & 25 & 1.5 & $97.7$ & $6.25$  \\
                             &     & 08~36~22.80 & $+$33~27~38.8 \\

SBS\,0927+493     & LINER & 09~31~06.76 & $+$49~04~47.5 & 10167  & 02-01-2006 & 37 & 2.3  & $195$ & $12.5$ \\
                  &       & 09~31~06.77 & $+$49~04~47.2            \\

UGC\,09618\,NED02     & LINER & 14~57~00.68 & $+$24~37~02.7 & 10094  & 01-15-2005 & 26 & 1.4 & $195$ & $12.5$ \\
                      &       & 14~57~00.67 & $+$24~37~02.9            \\

NGC\,6264       & Sy2 & 16~57~16.12 & $+$27~50~58.7 & 10177  & 01-27-2005 & 23 & 1.3 & $195$ & $12.5$ \\
                &     & 16~57~16.13 & $+$27~50~58.6 \\

\enddata

\tablenotetext{(a)}{Activity type from the NED.}

\tablenotetext{(b)}{{\it First line:} optical positions from the NED with
uncertainties of $\pm0\rlap{.}''5$ (SBS\,0927+493) or $\pm1\rlap{.}''3$. {\it
Second line:} maser positions measured with a VLA snapshot, providing typical
uncertainties of $\pm0\rlap{.}''3$. To establish the magnitude of systematic
uncertainties in position, we imaged disjoint segments of the VLA data and
confirmed that they yield consistent maser positions (that is, within
$0\rlap{.}''3$). }

\tablenotetext{(c)}{Optical heliocentric systemic velocity from the NED.}

\tablenotetext{(d)}{Average system temperature.}

\tablenotetext{(e)}{Rms noise in a $24.4$\,kHz spectral channel, corrected for
atmospheric opacity (typically from $0.02$ to $0.07$) and for the dependence of
antenna gain on elevation.}

\tablenotetext{(f)}{VLA channel width.}

\tablenotetext{(g)}{VLA bandwidth; the VLA band was in each case centered on the
strongest maser line.}

\end{deluxetable}

\clearpage

\begin{deluxetable}{llllrrrrl}
\tablewidth{0pt}
\tablecaption{Known Water Maser Sources Reobserved with the Green Bank
Telescope.\label{TableOther}} \tablehead{
     \colhead{Galaxy}              &
     \colhead{Type\tablenotemark{(a)}}              &
     \colhead{$\alpha_{2000}$\tablenotemark{(b)}}      &
     \colhead{$\delta_{2000}$\tablenotemark{(b)}}   &
     \colhead{v$_{sys}$\tablenotemark{(c)}}    &
     \colhead{Date}           &
     \colhead{$T_{sys}$\tablenotemark{(d)}}      &
     \colhead{$1\sigma$\tablenotemark{(e)}} \\
     \colhead{}             &
     \colhead{}             &
     \colhead{(hh~mm~ss)}     &
     \colhead{(dd~mm~ss)}     &
     \colhead{(km s$^{-1}$)} &
     \colhead{}             &
     \colhead{(K)} &
     \colhead{(mJy)} &
}

\startdata

IC\,0184        & Sy2 & 01~59~51.23 & $-$06~50~25.4 & 5382  & 07-03-2005 & 71 & 8.9  \\
VII\,ZW\,073    & Sy2 & 06~30~25.54 & $+$63~40~41.3 & 12391 & 01-01-2005 & 35 & 2.6 \\
MRK\,0034       & Sy2 & 10~34~08.592& $+$60~01~52.01& 15140 & 01-18-2005 & 21 & 0.98 \\
NGC\,3393       & Sy2 & 10~48~23.45 & $-$25~09~43.6 & 3750  & 01-15-2005 & 37 & 1.9 \\
NGC\,5495       & Sy2 & 14~12~23.35 & $-$27~06~29.2 & 6737  & 02-09-2006 & 40 & 2.4 \\
AM\,2158-380\,NED02  & Sy2 & 22~01~17.10 & $-$37~46~23.0 & 9983 & 07-03-2005 & 99 & 22 \\
\enddata

\tablenotetext{(a)}{Activity type from the NED.}

\tablenotetext{(b)}{Positions from Kondratko et al. 2006 ($\sigma=0\rlap{.}''3$)
except for MRK\,0034, which is from the NED ($\sigma=0\rlap{.}''3$).}

\tablenotetext{(c)}{Optical heliocentric systemic velocity from the NED.}

\tablenotetext{(d)}{Average system temperature.}

\tablenotetext{(e)}{Rms noise in a $24.4$\,kHz spectral channel, corrected for
atmospheric opacity (typically from $0.02$ to $0.07$) and for the dependence of
antenna gain on elevation.}

\end{deluxetable}

\begin{deluxetable}{lrccllc@{\hspace{3pt}}lc}
\tablewidth{6.5in} \tablecaption{Water Maser and Unabsorbed $2-10$\,keV
Luminosities.\label{XrayTable}} \tablehead{
     \colhead{Galaxy}              &
     \colhead{D\tablenotemark{(a)}}              &
     \colhead{$\mbox{log}\,L_{\mbox{\small H$_2$O}}$\tablenotemark{(b)}}              &
     \colhead{$\mbox{log}\,L_{\mbox{\small 2-10}}$\tablenotemark{(c)}}    &
     \colhead{Telescopes\tablenotemark{(d)}}    &
     \colhead{Ref\tablenotemark{(e)}}           &
     \multicolumn{2}{c}{Type\tablenotemark{(f)}}      &
     \colhead{Ref\tablenotemark{(g)}} \\
     \colhead{}             &
     \colhead{(Mpc)}             &
     \colhead{($L_{\odot}$)}     &
     \colhead{($L_{\odot}$)}     &
     \colhead{}       &
     \colhead{}       &
     \colhead{}        \\
}

\startdata

NGC\,1068    &    14  &  1.7  &  $>$9.6    & ABR & 1       & D & C & A   \\
Circinus     &   4.0  &  1.3  &  7.5-8.6   & B   & 2-4     & D & C & B    \\
NGC\,4945    &   4.0  &  1.7  &  9.1-9.6   & BCR & 5,6     & D & C & C \\
M\,51        &    10  & -0.2  &  7.7-8.4   & BG  & 7,8     & X & C & D \\
NGC\,3079    &    16  &  2.5  &  8.4-9.4   & B   & 9       & D & C & E \\
NGC\,1386    &    17  &  2.1  &  8.4-9.0   & BCX & 10,11   & D & C & F \\
NGC\,3393    &    50  &  2.4  &  8.5-10.2  & BX  & 10,12,13& d & C & G \\
IC\,2560     &    35  &  2.0  &  8.3-9.2   & AC  & 14,15   & D & C & H \\
NGC\,4051    &    10  &  0.3  &  7.6-8.5   & CBRX& 16-20   & X & c & I \\
NGC\,4388    &    34  &  1.1  &  8.7-9.5   & ABCRSX & 21   & d & c & J \\
3C\,403      &   235  &  3.3  &  9.6-10.3  & C   & 22      & X & c & K \\
NGC\,4258    &   7.2  &  1.9  &  7.0-7.7   & ABCX& 23      & D & c & L \\
NGC\,6240    &    98  &  1.6  &  9.8-10.8  & ABR & 24,25   & X & C & M \\
MRK\,266     &   110  &  1.5  &  9.3-10.3  & B   & 26      & X & C & J \\
MRK\,3       &    54  &  1.0  &  9.6-10.4  & BCGR& 27-32   & X & C & J \\
NGC\,5643    &    16  &  1.3  &  7.6-9.0   & BX  & 13,33   & X & C?& G \\
NGC\,5347    &    31  &  1.5  &  8.1-9.1   & A   & 14      & X & C & N \\
MRK\,1066    &    48  &  1.5  &  9.0-9.2   & A   & 34      & X & C & J  \\
ESO\,103-G\,035&  53  &  2.6  &  9.1-9.9   &ABEGR& 35-38   & X & c & N\\
NGC\,6300    &    15  &  0.34 &  8.1-8.5   & BRX & 35,39-41,67& X & c & G \\
NGC\,5506    &    25  &  1.7  &  8.9-9.4   &ABCGRX& 35,42-48&X & c & N \\
NGC\,1052    &    17  &  2.1  &  7.5-8.0   & ABC & 49-55   & J & c & O \\
MRK\,348     &    62  &  2.6  &  8.9-10.4  & AG  & 32,56,57& J & c & P \\
MRK\,1210    &    54  &  1.9  &  9.3-10.3  & AB  & 58,59   & d & C?& T \\
NGC\,2639    &    44  &  1.4  &  7.0-8.1   & A   & 52,59-61& X & c & N \\
UGC\,5101    &   157  &  3.2  &  8.8-9.8   & C   & 62      & X & C & R \\
NGC\,2273    &    25  &  0.8  &  8.1-9.3   & BX  & 10,13   & X & C & R \\
NGC\,2782    &    34  &  1.1  &  7.6-9.1   & C   & 63      & X & C & J \\
ARP\,299     &    42  &  2.1  &  8.2-9.4   & BCX & 64-66   & X & C & S \\
NGC\,5728    &    37  &  1.9  &  8.7-10.2  & C   & 63      & d & C?& J \\

\enddata

\tiny\tablenotetext{(a)}{Distances adopted from \cite{Henkel2005}.}

\tablenotetext{(b)}{Total isotropic water maser luminosities from
\cite{Henkel2005} except for NGC\,3393, NGC\,5643, NGC\,6300, NGC\,1068
(Kondratko et al. 2006), NGC\,3079 \citep{Kondratko2005}, UGC\,5101, and
NGC\,2273 \citep{Zhang2006}.}

\tablenotetext{(c)}{Unabsorbed $2-10$\,keV luminosity. In cases where this is not
described in the literature, the unabsorbed $2-10$\,keV luminosity has been
inferred from other reported parameters (flux, photon index, and the neutral
hydrogen column density) using WebPIMMS. The ranges quoted here reflect results
from multiple observations, alternative source models, and/or the uncertainty in
the reflection efficiency (in case of Compton-thick sources only; see note
$\dagger$ below). To account for source variability in case of 3C\,403, where
multiple observations and alternative models were not available, we adopted the
fractional uncertainty averaged over all Compton-thin systems in our sample. Only
observations sensitive to energies above $10$\,keV were considered for
Compton-thick sources, with the exception of NGC\,1386, IC\,2560, NGC\,5347,
MRK\,1066, UGC\,5101, NGC\,2782, and NGC\,5728 for which hard X-ray data were not
available.}

\tablenotetext{(d)}{X-ray telescopes used in luminosity measurements: A$=$ASCA,
B$=$BeppoSAX, C$=$CXO, E$=$EXOSAT, G$=$GINGA, R$=$RXTE, S$=$SL2-XRT, X$=$XMM.}

\tablenotetext{(e)}{References for the $2-10$\,keV luminosity: [1]
\cite{Colbert2002} but also see \cite{Comastri2004} and references therein, [2]
Guainazzi et al. (1999)\nocite{Guainazzi1999}, [3] Matt et al.
(1999)\nocite{Matt1999}, [4] \cite{Smith2001}, [5] \cite{Guainazzi2000}, [6]
\cite{Done2003}, [7] \cite{Fukazawa2001}, [8] \cite{Makishima1990}, [9]
\cite{Iyomoto2001}, [10] \cite{Guainazzi2005}, [11] \cite{Comastri2004}, [12]
\cite{Salvati1997}, [13] \cite{Maiolino1998}\tablenotemark{\dagger}, [14]
\cite{Risaliti1999}\tablenotemark{\dagger}, [15] \cite{Iwasawa2002}, [16]
\cite{Pounds2004}, [17] \cite{Uttley2004}, [18] \cite{McHardy2004}, [19]
\cite{Uttley2003}, [20] \cite{Uttley1999}, [21] \cite{Elvis2004}, [22]
\cite{Kraft2005}, [23] \cite{Fruscione2005}, [24] \cite{Ikebe2000}, [25] Vignati
et al. (1999)\nocite{Vignati1999}, [26]
\cite{Risaliti2000}\tablenotemark{\dagger}, [27] \cite{Lutz2004}, [28]
\cite{Matt2000}, [29] \cite{Sako2000}, [30] Capi et al. (1999)\nocite{Capi1999},
[31] \cite{Georgantopoulos1999}, [32] \cite{Smith1996}, [33]
\cite{Guainazzi2004}, [34] \cite{Levenson2001}, [35] \cite{Risaliti2002}, [36]
\cite{Wilkes2001}, [37] \cite{Akylas2001}, [38] \cite{Georgantopoulos2001}, [39]
\cite{Maddox2002}, [40] \cite{Guainazzi2002}, [41] \cite{Leighly1999}, [42]
\cite{ONeill2005}, [43] \cite{Bianchi2003}, [44] \cite{Perola2002}, [45]
\cite{Matt2001}, [46] \cite{Lamer2000}, [47] \cite{Wang1999}, [48]
\cite{Nandra1994}, [49] \cite{Kadler2004}, [50] \cite{Satyapal2004}, [51]
\cite{Ueda2001}, [52] \cite{Terashima2000}, [53] \cite{Guainazzi2000b}, [54]
\cite{Weaver1999}, [55] \cite{Guainazzi1999b}, [56] \cite{Akylas2002}, [57]
\cite{Warwick1989}, [58] \cite{Ohno2004}, [59]
\cite{Bassani1999}\tablenotemark{\dagger}, [60] \cite{Terashima2002}, [61] Wilson
et al. (1998)\nocite{Wilson1998}, [62] Assuming that the source is Compton-thick
based on Armus et al. (2004)\nocite{Armus2004} and using the measured flux from
\cite{Ptak2003}\tablenotemark{\dagger}, [63]
\cite{Zhang2006}\tablenotemark{\dagger}, [64] \cite{DellaCeca2002}, [65]
\cite{Ballo2004}\tablenotemark{\dagger}, [66]
\cite{Zezas2003}\tablenotemark{\dagger}, [67] \cite{Matsumoto2004}. }

\tablenotetext{(f)}{Two letter code describing (1) maser type and (2) magnitude
of X-ray absorption column. Maser type: d$=$high-velocity water maser system,
i.e., a source where origin of emission in an edge-on disk is suggested by
spectral data only; D$=$high-velocity water maser systems, where hypothesized
disk origin has been reinforced by an analysis of VLBI data; J$=$maser source
associated with a radio jet; X$=$non-high-velocity system. Obscuration:
C$=$Compton-thick; c$=$Compton-thin. A question mark after obscuration type
indicates controversy in analysis of X-ray data reported in the literature.}

\tablenotetext{(g)}{References for the maser type: [A] \cite{Greenhill1997c}, [B]
\cite{Greenhill2003}, [C] \cite{Greenhill1997}, [D] \cite{Hagiwara2001}, [E]
\cite{Kondratko2005}, [F] \cite{Braatz1997AAS}, [G] Kondratko et al. 2006, [H]
\cite{Ishihara2001}, [I] \cite{Hagiwara2003}, [J] \cite{Braatz2004}, [K]
\cite{Tarchi2003}, [L] \cite{Miyoshi1995}, [M] \cite{Hagiwara2003b}, [N]
\cite{Braatz1996}, [O] \cite{Claussen1998}, [P] \cite{Peck2003}, [R]
\cite{Zhang2006}, [S] \cite{Henkel2005} [T] Kondratko et al., in preparation.}

\tablenotetext{\dagger}{Following \cite{Comastri2004}, we assumed in case of
Compton-thick models for NGC\,5347, NGC\,5643, NGC\,3393, MRK\,266, MRK\,1210,
UGC\,5101, NGC\,2273, NGC\,2782, ARP\,299, and NGC\,5728 that the observed
$2-10$\,keV luminosity is $1-10\%$ of the unabsorbed $2-10$\,keV luminosity due
to reflection and scattering.}

\end{deluxetable}

\clearpage

\begin{deluxetable}{lc@{\hspace{3pt}}cc@{\hspace{3pt}}cc@{\hspace{3pt}}cc}
\tablewidth{0pt} \tablecaption{Correlation Coefficients and Power Law Indices
for $L_{2-10}$ vs. $L_{\mbox{\tiny H$_2$O}}$.\label{fits}} \tablehead{
     Population              &
     \multicolumn{2}{c}{$m$\tablenotemark{(a)}}      &
     \multicolumn{2}{c}{$\rho$\tablenotemark{(b)}}  &
     \multicolumn{2}{c}{Significance\tablenotemark{(c)}}  &
     N\tablenotemark{(d)}  \\

}

\startdata

all systems                     & $0.4\pm0.1$ & ($0.4\pm0.1$)    & $0.4\pm0.1$ & ($0.3\pm0.1$) & $7\pm 7$ & ($20\pm 10$) & 29     \\

all systems\tablenotemark{\dagger}& $0.5\pm0.1$ & ($0.5\pm0.1$)   & $0.5\pm0.1$ & ($0.4\pm0.1$) &  $3\pm 4$ & ($10\pm 10$) & 20 \\

unclassified and jet masers   & $0.5\pm0.1$ & ($0.4\pm0.1$)   & $0.5\pm0.1$ & ($0.4\pm0.1$) & $5\pm 6$ & ($20\pm 10$) & 19      \\

high-velocity                 & $0.4\pm0.3$ & ($1.0\pm0.5$)   & $0.1\pm0.2$ & ($0.2\pm0.2$) & $70\pm 20$ & ($60\pm 20$) & 10     \\

Compton-thick                 & $0.3\pm0.1$ & ($0.2\pm0.2$)   & $0.3\pm0.1$ & ($0.1\pm0.1$) & $30\pm 20$ & ($60\pm 20$) & 19   \\

Compton-thin                  & $0.6\pm0.1$ & ($0.5\pm0.2$)    & $0.5\pm0.1$ & ($0.3\pm0.1$) & $20\pm 10$ & ($40\pm 20$) & 10     \\

\enddata

\tablenotetext{(a)}{Index ($L_{2-10}\propto L^m_{\tiny\mbox{H$_2$O}}$) obtained
from a Monte Carlo simulation where the probability distribution function for
each data point in Fig. \ref{Xray} is modelled in linear luminosity space by a
Gaussian with mean and standard deviation corresponding to the plotted symbols
and the error bars in Fig. \ref{Xray}, respectively. The parentheses enclose the
result for the same population but with the leftmost and the rightmost data
points in Fig. \ref{Xray} removed.}

\tablenotetext{(b)}{Spearman correlation coefficient and uncertainty from the
Monte Carlo simulation described in (a). Alternative use of a Pearson correlation
coefficient changes the results negligibly but assumes in the calculation of the
significance level that the two variables are normally distributed, which is most
likely not the case here. Quantities in parentheses are as in (a).}

\tablenotetext{(c)}{Two-sided significance level (expressed as a percentage),
i.e., probability that a random data set of the same size would yield a larger
magnitude of the correlation coefficient. Quantities in parentheses are as in
(a). We have confirmed the reported significance levels with numerous
($\sim10^5$) realizations of the permutation test, whereas the null hypothesis of
no correlation is assumed and the water luminosity data points are permuted among
the x-ray luminosity data points \citep[e.g.,][]{Wall2003}.}

\tablenotetext{(d)}{Number of sources. The lower limit on the X-ray luminosity of
NGC\,1068 was not used in the fit.}

\tablenotetext{\dagger}{Galaxies with multiple nuclei (NGC\,6240, MRK\,266,
MRK\,1066, ARP\,299), jet masers (MRK\,348, NGC\,1052), and LINER systems
(NGC\,3079, NGC\,4258, NGC\,6240, NGC\,1052, UGC\,5101) have been removed from
the sample. See the text for details.}

\end{deluxetable}

\clearpage

\end{document}